\documentclass[aps,pre,reprint,superscriptaddress,nofootinbib]{revtex4-2}

\usepackage[T1]{fontenc}
\usepackage[utf8]{inputenc}
\usepackage{lmodern}
\usepackage{microtype}
\usepackage{amsmath,amssymb}
\usepackage{hyperref}
\usepackage{xcolor}

\hypersetup{
  colorlinks=true,
  linkcolor=black,
  citecolor=black,
  urlcolor=blue,
  pdftitle={When higher-order interactions matter: reducibility, parsimony, and microscopic organization}
}

\begin{document}

\title{When higher-order interactions matter: reducibility, parsimony, and microscopic organization}

\author{Alex Arenas}
\email{alexandre.arenas@urv.cat}
\affiliation{Universitat Rovira i Virgili, Tarragona, Spain}

\author{Federico Battiston}
\email{fbattiston@luiss.it}
\affiliation{Luiss University, Rome, Italy}
\affiliation{Central European University, Vienna, Austria}

\author{Andrea Gabrielli}
\email{andrea.gabrielli@uniroma3.it}
\affiliation{Department ICITA, Universit\`a degli Studi ``Roma Tre'', Rome, Italy}
\affiliation{``Enrico Fermi'' Research Center - CREF, Rome, Italy}

\date{\today}

\begin{abstract}
The current debate on higher-order interactions raises a fundamental question: when can systems organized through group interactions be faithfully represented within a graph-based formalism? Recent results show that graph descriptions can reproduce higher-order dynamics exactly or preserve selected macroscopic observables. Yet reducibility does not necessarily imply simplification, as information removed from the interaction structure may reappear as complexity in the effective dynamics, while structural features such as nestedness, heterogeneity, and cross-order correlations may be obscured by the reduction. We argue that formal representability alone is insufficient as a criterion for model choice. Both ``simpler'' and ``equivalent'' are question-dependent notions: parsimony must be assessed for the complete structure--dynamics model, and adequacy must be defined relative to the observables, scales, and regimes required by the scientific question. A reduced model may therefore be fully adequate for locating a phase transition or identifying a universality class while being inadequate for reproducing microscopic trajectories, transient dynamics, or structure--function relations. When group structure carries relevant microscopic information, higher-order representations may remain the most direct and interpretable description and, under an explicit model-selection or complexity criterion, may also provide the most parsimonious one, as well as the natural framework for understanding how microscopic organization gives rise to macroscopic behavior and functionality.
\end{abstract}

\maketitle

\section*{Introduction}

As a wide variety of empirical systems are naturally described by interactions involving groups rather than only pairs, higher-order interactions have emerged as a powerful modeling framework and an active research topic in network science \citep{battiston2020networks}. Hypergraphs, simplicial complexes, and related formalisms represent such non-dyadic interactions explicitly, providing a framework to describe the organization of complex systems beyond pairwise connections \citep{battiston2021physics,bianconi2021higher,bick2023what}. They have been used to study contagion~\citep{iacopini2019simplicial}, synchronization~\citep{skardal2020higher}, cooperation~\citep{alvarezrodriguez2021evolutionary}, and other collective processes, showing that the organization of interactions at the group level can profoundly affect macroscopic behavior~\citep{battiston2026collective}.

The growing use of higher-order representations has also raised the question of when they are needed. Recent results show that some dynamics defined on higher-order structures can be represented exactly by pairwise models, while in other cases pairwise mechanisms can reproduce only selected macroscopic behavior commonly associated with higher-order interactions \citep{meloni2026higher,llabres2026reducibility,tan2026replicating}. Moreover, nonlinear dynamics on graphs can sometimes be represented as linear dynamics on a richer higher-order state space \citep{lacasa2026equivalence}. At the same time, a graph does not, by itself, restrict the dynamical rule to pairwise interactions: multivariate interaction functions may act over graph neighborhoods, so topological representation and dynamical interaction order should be kept conceptually distinct~\citep{peixoto2026graphs}.

Here we argue that formal expressivity and model choice are different questions. Our concern, however, is not which formalism is maximally expressive, but which representation most directly connects the organization of the system to the scientific question being asked. A pairwise representation may exist without providing a ``simpler'' description, since information removed from the interaction structure can reappear in the effective dynamics, while meaningful structural organization can be lost in the reduction. In this Perspective, we examine recent reducibility results to clarify when higher-order interactions can be removed without loss, when reduction merely shifts complexity elsewhere in the model, and when keeping group structure explicit provides the more direct and parsimonious representation of a complex system.

\section*{Formal expressivity is not the same as scientific representation}

A mathematical representation should not be evaluated only by the set of behaviors it can generate. Scientific models are also judged by how directly their mathematical objects correspond to the entities and interactions that are measured or hypothesized in the system. If the elementary observation is a group event---for example, a collaboration, a multi-species encounter, or a multi-component reaction---then a hyperedge may be the most direct representation of that datum. Projecting the event onto pairs is certainly possible, but it changes what is treated as primitive information. Here and below, by a pairwise projection we mean an ordinary dyadic projection; this should not be conflated with the general expressive capacity of graph-based dynamical models.

This distinction is familiar throughout network science. Over time, the field has progressively enriched its microscopic representations through directed~\citep{wasserman1977random}, weighted~\citep{barrat2004architecture}, signed~\citep{cartwright1956structural}, temporal~\citep{holme2012temporal}, multilayer~\citep{kivela2014multilayer}, and adaptive networks~\citep{gross2008adaptive}. These extensions were not introduced because a sufficiently complicated dynamical rule on a simpler graph could never reproduce their macroscopic consequences. They were introduced because directions, weights, signs, times, and interaction types can be part of the measured organization of a system. Higher-order representations follow the same modeling logic \citep{battiston2020networks}.

The value of an explicit representation is therefore not only that it may make some calculations easier. It also determines which questions are natural to ask. A pairwise projection emphasizes dyadic connectivity. A higher-order representation makes group size~\citep{patania2017shape}, nestedness~\citep{lotito2022higher}, cross-order correlations~\citep{gallo2024higher}, and other structural observables explicit. These quantities may themselves be the scientific objects of interest, independently of whether a different formalism can reproduce a selected dynamical observable.

\section*{Model reduction and scientific questions}

Whether a reduced representation is useful depends not only on whether it can reproduce the original model, but also on what one means by simpler and by equivalent. Simplicity may refer to the mathematical form of the model, its computational cost, dimensionality, number of parameters, the amount of information required for its specification, or the ease with which its parameters can be inferred from data. In some settings, simplification can be defined precisely, as in exact rank reduction, where a network model is represented in a lower-dimensional space while preserving its exact description \citep{valdano2019exact}. More generally, however, simplifying the structural representation need not simplify the complete model, as discussed below.

Similarly, whether two models should be regarded as equivalent depends on the scientific question being addressed. A reduced description need not reproduce every feature of the original model to be useful: for some questions microscopic trajectories matter, whereas for others reproducing a phase transition or its critical behavior may be sufficient. The relevant notion of equivalence must therefore be specified together with the property the model is intended to describe.

This distinction is familiar in statistical physics. The Ising model is a highly simplified description of real magnetic systems, yet it captures the essential physics of the ferromagnetic phase transition and, through universality, also describes the critical behavior of systems with very different microscopic realizations. Its usefulness does not derive from microscopic fidelity, but from preserving the features relevant to the question being asked  and from its ability to explain the observed universality of behavior of systems that, at first sight, appear very different. The same principle applies when comparing pairwise and higher-order representations: the appropriate level of reduction depends on which properties of the original system one aims to preserve and over which range of parameters.

\section*{Exact reducibility and the cost of reduction}

The analysis of Llabr\'es \emph{et al.} provides a particularly clean way to separate formal equivalence from model choice \citep{llabres2026reducibility}. They consider binary-state social-impact models in which a focal node first selects one of its incident hyperedges and then changes state with a probability $f(\phi_e)$ determined by the fraction $\phi_e$ of opposite-state nodes in that group. For node-update dynamics and social-impact functions satisfying the absorbing-state conditions of their construction, they derive a pairwise imitation process on the projected network whose transition probabilities are \emph{exactly} the same at the microscopic level. This is a strong reducibility theorem: it does not merely reproduce a mean-field equation or {capture features of a phase transition, such as critical exponents and universality classes}.

The mechanism of the reduction is, however, as important as its existence. A nominal pairwise projection preserves neighborhood relations but loses explicit information about which nodes belong to which groups and, in particular, does not preserve hyperedge overlap. Exact equivalence is recovered by assigning directed effective weights $\omega_{ij}$ to projected links. Those weights are obtained by summing contributions from the hyperedges shared by $i$ and $j$ and therefore encode information about the original higher-order architecture. In the generic case they also depend on the local state composition of those hyperedges and hence on the instantaneous configuration of the system. The reduced object is consequently not simply a static graph with ordinary pairwise couplings. It is, in general, a state-dependent weighted network whose dynamics contains the information that was explicit in the hypergraph.

More precisely, the same paper identifies a regime in which reduction is both rigorous and genuinely simplifying. For a linear social-impact function, the effective weights become state-independent and depend only on structural quantities such as hyperdegrees and hyperedge orders. Moreover, for the random hypergraph ensembles studied, the macroscopic ordering dynamics is independent of the heterogeneity of these weights and coincides with that of the standard voter model on the unweighted projected network. In this regime, the pairwise representation is not only formally equivalent but also operationally parsimonious.

The nonlinear case is different and explicitly shows how nontrivial the reducibility problem is. The exact microscopic reduction still exists, but the projected weights evolve with the state configuration. Thus, the paper itself distinguishes three logically different statements: exact microscopic reducibility, approximate macroscopic reducibility and its range of applicability, and practical minimal-model adequacy at a specified accuracy.

The boundary of the theorem is equally instructive. Llabr\'es \emph{et al.} note that hyperedge-update rules, in which all nodes in a selected group update simultaneously, cannot by construction be mapped onto a microscopically equivalent pairwise process that updates only one node per interaction. Macroscopic equivalence may still occur in special cases, but it becomes a separate question. This is precisely the kind of conditional statement that a productive higher-order versus pairwise debate should aim for: not whether one formalism wins universally, but which classes of mechanisms are reducible, to what representation, at what informational cost, and with respect to which scientific question.

\section*{The structure--dynamics duality and where to place complexity}

A useful way to formulate this trade-off is to regard a model $\mathcal{M}$ as a pair
\begin{equation}
    \mathcal{M}=(\mathcal{S},\mathcal{F}),
\end{equation}
where $\mathcal{S}$ denotes the  topological structure of interactions and $\mathcal{F}$ the dynamical rules. A reduction from a higher-order model to a graph generally changes both components. Information removed from $\mathcal{S}$ may reappear in $\mathcal{F}$ through multivariate functions, state-dependent edge weights, auxiliary variables, or additional layers. The construction of Llabr\'es \emph{et al.} discussed above provides a concrete example of this transfer of information from structure to dynamics \citep{llabres2026reducibility}. Here, reduction refers to a mapping from a specified model or representation to a target description, rather than to an ordering of entire classes of mathematical models. The opposite  direction is also possible, since nonlinear graph dynamics can sometimes be represented as linear dynamics on a richer higher-order structure \citep{lacasa2026equivalence}  or in a higher-dimensional space \citep{brunton2022, diantonio2026}. In a related direction, Neuh\"auser \emph{et al.} distinguish topological from dynamical order and define the effective order as the minimum interaction order required to reproduce the observed dynamics \citep{neuhauser2024learning}.

Parsimony should therefore be assessed for the complete model rather than for its structural substrate alone. It can be made operational through Bayesian model selection, the minimum description length (MDL) principle, or other explicit complexity criteria~\citep{peixoto2026graphs}. At the same time, structural description length, dynamical complexity, inference requirements, and computational cost need not induce the same ordering of representations. The relevant criterion should therefore be stated explicitly whenever one representation is claimed to be more parsimonious than another. 

An ordinary dyadic projection may be structurally simpler than an explicit higher-order representation, but the resulting model need not be simpler if the reduction requires a high-dimensional or strongly state-dependent dynamical rule. Conversely, an explicit higher-order structure can keep the dynamics local and homogeneous. Context-dependent spreading provides a concrete example. Information about the group in which an interaction takes place can be kept explicitly in the structure or incorporated into the rule governing an otherwise pairwise interaction \citep{burgio2023spreading}. Different representations can therefore encode the same information in different parts of the model. A useful reduction should remove complexity from the model as a whole, rather than simply transfer it from structure to dynamics, while preserving the system features one aims to study.

This distinction also matters for inference. In the nonlinear social-impact reduction, the effective pairwise weights require knowledge of the composition of the original hyperedges and change as the state configuration evolves. The pairwise representation can therefore be exact while being less direct to infer from data recorded as group events. When the induced weights are static, or when the observables of interest are insensitive to them, the reduced description may instead be preferable. Formal equivalence alone does not decide between these cases. The relevant comparison also includes how much information each representation requires and how directly its parameters can be obtained from data.

\section*{Equivalence is relative to what must be preserved}

Statements of equivalence require specifying what is preserved and over which region of parameter space. A reduction may preserve microscopic transition probabilities, trajectories, stationary distributions, mean-field equations, bifurcation structure, universality classes, or only selected observables. It may do so globally, only asymptotically, only near a critical point, or only within a prescribed error tolerance. These are different levels of equivalence and should not be conflated.

The distinction is already visible in the results discussed above. Llabr\'es \emph{et al.} establish exact microscopic equivalence through a weighted pairwise representation, while the unweighted projection is assessed separately through its ability to reproduce selected macroscopic observables \citep{llabres2026reducibility}. Tan \emph{et al.} provide a complementary numerical example for epidemic dynamics: a static normalization of the pairwise infection parameter can closely reproduce higher-order endemic steady states, but not the transient trajectory; matching both transient and stationary behavior requires a time-dependent effective infection parameter, and the approximation weakens for heterogeneous hypergraph topologies \citep{tan2026replicating}. Meloni \emph{et al.} instead address coarse-grained critical behavior and universality for a class of contagion processes \citep{meloni2026higher}. Peixoto \emph{et al.} consider a different question again, namely the representational capacity of graph-based models when sufficiently general dynamical functions are allowed \citep{peixoto2026graphs}. These results establish different forms of equivalence rather than competing versions of the same statement.

The distinction matters because the same macroscopic behavior can arise from different microscopic organizations. Agreement at the level of a mean-field equation, a phase transition, or a universality class does not imply agreement in local correlations, group composition, or mesoscopic organization. Conversely, an exact microscopic mapping for one class of dynamics does not imply that other higher-order processes admit reductions with the same properties. A claim of reducibility should therefore specify what survives the reduction and what does not.  In other words, it must always be defined in relation to the precise scientific questions posed about the studied system.

\section*{Beyond mean field: the role of microscopic organization}

The distinction becomes particularly important beyond homogeneous or mean-field settings. Much of the current higher-order literature is concerned not simply with the presence of interactions of order three or larger, but with how those interactions are organized. Degree heterogeneity~\citep{landry2020effect}, cross-order correlations~\citep{gallo2024higher}, overlap, and nestedness~\citep{lotito2022higher} provide examples of structural information that is ubiquitous in empirical systems but invisible or strongly transformed under many pairwise projections.

Analytical closures that retain group correlations make this point explicit. Clique-cover and triadic approximations show that higher-order correlations and overlap between pairwise and three-body interactions can shift contagion thresholds and outbreak sizes beyond standard mean-field predictions \citep{burgio2021network,burgio2024triadic}. Approximate-master-equation approaches reach a similar conclusion for heterogeneous and adaptive hypergraphs \citep{stonge2022influential,burgio2025characteristic}.

The organization of higher-order interactions can also determine qualitative features of collective dynamics. Zhang \emph{et al.} showed that hypergraphs and simplicial complexes can produce markedly different synchronization behavior, with the outcome controlled by higher-order degree heterogeneity and cross-order correlations \citep{zhang2023higher}. More recently, Malizia \emph{et al.} showed that nested hyperedges can simultaneously promote the onset of collective behavior and suppress explosive transitions, with the same mechanism emerging across contagion, Ising, and Kuramoto dynamics \citep{malizia2026nested}. These results show that interaction order alone does not characterize the microscopic organization controlling collective behavior. At the same time, the occurrence of an abrupt transition should not, by itself, be interpreted as evidence for a particular higher-order representation: the same macroscopic behavior can arise in graph-based models with suitable multivariate interaction rules~\citep{peixoto2026graphs}. Macroscopic phenomenology alone therefore need not identify the underlying microscopic organization.

A reduction that preserves a macroscopic observable  or some dynamical or statistical features within a specific parameter region can therefore remove structural information that controls other aspects of the dynamics. Whether this matters depends on what the reduced model is expected to reproduce.  If, in this process, one realizes that nestedness, heterogeneity, or correlations across interaction orders contribute to the mechanism under study, or reflect empirical observations, preserving them explicitly may be essential even when some macroscopic behavior survives projection.

\section*{Reduction is part of higher-order network science}

The computational cost of higher-order representations is a genuine limitation. They require more memory, more complex algorithms, and often larger parameter spaces. This is exactly why reduction should be treated as a central problem rather than as an argument that higher-order representations are unnecessary.

Reduction is a broader problem in network science. Thibeault \emph{et al.} advanced a low-rank hypothesis for complex systems, showing how some high-dimensional network dynamics can be reduced to lower-dimensional descriptions \citep{thibeault2024low}. Related questions have also been addressed for richer network representations. De Domenico \emph{et al.}, for instance, investigated the reducibility of multilayer networks, identifying when layers can be aggregated while retaining the information carried by the multilayer representation \citep{dedomenico2015structural}. More recently, Lacasa introduced reduction methods for temporal networks that enable their compression and projection onto low-dimensional representations \citep{lacasa2026fluid}.

Recent work extends this direction to higher-order networks. Kirkley \emph{et al.} introduced information-theoretic frameworks for identifying entire interaction orders \citep{kirkley2025structural} or individual hyperedges \citep{kirkley2026hypergraph} that are structurally redundant and can therefore be eliminated while retaining essential higher-order structural information. Neuh\"auser \emph{et al.} introduced the notion of effective order and developed a method to infer the minimum interaction order required to reproduce the observed dynamics \citep{neuhauser2024learning}. Lucas \emph{et al.} proposed a dynamics-informed reducibility framework that asks when higher-order structure can be compressed while preserving dynamical information \citep{lucas2026reducibility}. A complementary perspective was recently introduced by Rommens \emph{et al.}, who defined the representational complexity of a dynamical system in terms of the information required to specify it through a given interaction structure and dynamical rule \citep{rommens2026informational}. Their framework characterizes how this information can be distributed between the structural and dynamical components of a representation, providing an information-theoretic measure for comparing alternative structure--dynamics descriptions of the same system. These approaches make reduction itself a quantitative problem and show that different systems can support different degrees and forms of reduction.

Taken together, these approaches point toward a broader program for higher-order network science. Rather than retaining every observed higher-order interaction or reducing every system to pairwise interactions, one can ask which interactions, interaction orders, or structural features can be removed without losing the information that the model is intended to preserve. Developing a theory of when higher-order interactions can be removed  should therefore be seen as an integral part and a fundamental step of higher-order network science. The central question is which higher-order features of an original model can be reduced or suppressed so that the aspects of interest of the investigated real system are preserved.

\section*{From expressivity to modeling practice}

The higher-order versus pairwise debate ultimately concerns more than the expressive power of different formalisms. A central goal of network science has always been to understand how microscopic organization gives rise to macroscopic behavior. Network representations make this organization measurable and allow us to investigate how specific structural features shape collective phenomena. The relevant question is therefore not only whether a given behavior can be reproduced within a simpler formalism, but which representation most directly connects the organization observed in a system to the behavior we seek to understand. Model validity is consequently conditional on purpose: a reduction that is sufficient for detecting a transition or recovering a universality class need not be sufficient for predicting transients, local correlations, intervention responses, or microscopic mechanisms.

This perspective is not specific to higher-order interactions. Network science has progressively developed generalized representations to retain aspects of microscopic organization that are present in empirical systems, while reducibility has become a scientific question in its own right. Reducibility is therefore not a test of whether a generalized network framework is legitimate. It tells us how much of the information represented by that framework is required for a given purpose.

These considerations become particularly important when microscopic organization is inferred from empirical data. When group events are directly observed, their membership, sizes, nestedness, heterogeneity, and correlations across interaction orders characterize the empirical organization represented by the data. When higher-order structure is reconstructed from lower-order or temporally aggregated observations, these properties may also depend on the observation scale and reconstruction procedure. Whether such observables identify a particular microscopic interaction mechanism is a separate inferential question. Their dynamical consequences may sometimes be reproduced after projection, but this does not eliminate the scientific question of how the observed organization produces collective behavior. Representing that organization explicitly can therefore remain essential when the organization itself is part of what we seek to understand.

This leads to two complementary questions for higher-order network science. How does microscopic higher-order organization shape macroscopic behavior and functionality? And how much of that organization can be removed while preserving the information required for a given purpose? The first seeks the mechanisms connecting structure to collective phenomena. The second establishes which information is redundant and at what level of description. Together, they turn reducibility from an argument about the necessity of higher-order interactions into a tool for establishing their domain of relevance. The goal is not to declare higher-order interactions indispensable or dispensable, but to determine which aspects of microscopic organization must remain explicit for the scientific question at hand.

\begin{acknowledgments}
This work has been supported by Spanish Ministerio de Ciencia, Innovación y Universidades PID2024-158120NB-C21. A.A. also acknowledges the ICREA Academia program of Generalitat de Catalunya. F.B. acknowledges support from the Italian Fund for Science (Fondo Italiano per la Scienza, FIS) of the Italian Ministry of University and Research (MUR), FIS 2 call, through the project CODYBE (FIS-2023-02989; CUP B53C25003310001), funded under Decreto Direttoriale No. 1236 of 1 August 2023 and admitted to funding by D.D. Prot. No. I.0018353 of 19 November 2025. F.B. also acknowledges support from the Austrian Science Fund (FWF) through project STRHOINET (10.55776/PAT1052824).
\end{acknowledgments}

\end{document}